# Path-Sensitive Atomic Commit: Local Coordination Avoidance for Distributed Transactions

Tim Soethout[a,b], Tijs van der Storm[b,c], and Jurgen J. Vinju[b,d]

a ING Bank, Amsterdam, The Netherlands
b Centrum Wiskunde & Informatica, Amsterdam, The Netherlands
c University of Groningen, Groningen, The Netherlands
d Eindhoven University of Technology, Eindhoven, The Netherlands

**Abstract**  *Context* Concurrent objects with asynchronous messaging are an increasingly popular way to structure highly available, high performance, large-scale software systems. To ensure data-consistency and support synchronization between objects such systems often use distributed transactions with Two-Phase Locking (2PL) for concurrency control and Two-Phase commit (2PC) as atomic commitment protocol.
   *Inquiry* In highly available, high-throughput systems, such as large banking infrastructure, however, 2PL becomes a bottleneck when objects are highly contended, when an object is queuing a lot of messages because of locking.
   *Approach* In this paper we introduce Path-Sensitive Atomic Commit (PSAC) to address this situation. We start from message handlers (or methods), which are decorated with pre- and post-conditions, describing their guards and effect.
   *Knowledge* This allows the PSAC lock mechanism to check whether the effect of two incoming messages at the same time are independent, and to avoid locking if this is the case. As a result, more messages are directly accepted or rejected, and higher overall throughput is obtained.
   *Grounding* We have implemented PSAC for a state machine-based DSL called Rebel, on top of a runtime based on the Akka actor framework. Our performance evaluation shows that PSAC exhibits the same scalability and latency characteristics as standard 2PL/2PC, and obtains up to 1.8 times median higher throughput in congested scenarios.
   *Importance* We believe PSAC is a step towards enabling organizations to build scalable distributed applications, even if their consistency requirements are not embarrassingly parallel.

**ACM CCS 2012**

- **Information systems** → *Distributed database transactions*;
- **Software and its engineering** → *Domain specific languages*; State systems; Model-driven software engineering;
- **Applied computing** → Enterprise architectures; Event-driven architectures;

# The Art, Science, and Engineering of Programming

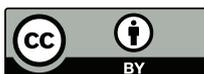

**Path-Sensitive Atomic Commit: Local Coordination Avoidance for Distributed Transactions**

# 1 Introduction

Structuring a software system as a collection of actively communicating objects is an increasingly popular architecture for large-scale, high performance, and high availability IT infrastructure. A common challenge in systems is to maintain high availability and consistency when communicating objects need to synchronize. This is particularly challenging in the context of large, scalable, highly available enterprise software. Our experience in the context of ING Bank[1] is that financial institutions deal with large and complex IT landscapes, consisting of many communicating software applications and components under high request loads, which need to synchronize to keep data consistent. These systems often perform operations that span multiple different applications and server nodes with consistency and durability guarantees.

A safe and well-known distributed transaction protocol to implement these distributed transactions, is Two-Phase Locking (2PL) [35] for isolation with Two-Phase Commit (2PC) [19] for atomicity. While this approach ensures consistency and serializability, it limits throughput in high-contention objects [5, 22, 27], since transactions have to wait on locks of other transactions on the same object. High-contention objects limit the throughput and latency of other objects they communicates with. Depending on the use case, this can be a problem. Examples with high-contention and strong consistency requirements are:

- tax bank accounts, involved in many money transfers and strict regulations on turnaround time;
- a video view counter on a popular video used for advertisement income calculations;
- cases with long-running transactions, where objects stay locked for long periods.

More general, applications with a long tail usage pattern, combined with strict performance and consistency requirements, will have high-contention objects.

This paper studies the performance of high-load strict 2PL/2PC in high- and low-contention use cases and introduces a novel concurrency mechanism named Path-Sensitive Atomic Commit (PSAC), which minimizes waiting in busy entities by exploiting high-level, functional knowledge about object behavior to reduce contention.

PSAC trades computing power for reduced waiting on locks, in order to achieve higher throughput than strict 2PL. By detecting whether two or more incoming requests have independent effects PSAC can start processing more requests in parallel than 2PL.

A request is independent of an already in-progress request if the acceptance or rejection of it is not influenced by whether the in-progress requests commit or abort. More details are discussed in section 3.

PSAC works under the assumptions that:

- all objects are state machine-based objects with clearly defined actions;
- the behavior of actions is defined by pre- and post-conditions on the local object, using first order logic with support for integer constraints, respectively describing their applicability and state effect;

---

[1] https://www.ing.com

- an object handles an action as an atomic step, checking the preconditions and applying its post-conditions;
- objects communicate by synchronized actions, which describe an atomic step of a group of actions on multiple objects;
- a group of actions is effectively a transaction among multiple objects.

Separating this functional specification of business objects from their implementation allows experimenting with different back-ends. In this case we have developed a code generator mapping high-level specifications, written in a state machine based domain specific language for financial products called Rebel [43, 44, 45], to an implementation based on the Akka actor framework, employing either 2PL/2PC or PSAC. The PSAC back-end then exploits the model's action pre- and post-conditions to detect independence of actions at run time.

Based on these two implementations we evaluate the performance of PSAC, and compare its performance in the same scenario to the standard of distributed transactions (ACID [21]), which is 2PL/2PC. Our results show that PSAC consistently outperforms 2PL/2PC in high-contention scenarios. Furthermore, PSAC retains the same scalability characteristics as 2PC, but does not guarantee serializability.

The contributions of this paper are as follows:

- We introduce PSAC, a novel concurrency mechanism that exploits semantics of operations to allow transactions to proceed in parallel if it can be detected that their effects are independent (section 3).
- We describe the implementation of PSAC based on Rebel, targeting the Akka actor framework, which provides the basis for our experimental setup (section 4).
- We evaluate the performance of both 2PL/2PC and PSAC, and show that PSAC outperforms 2PL/2PC in high-contention scenarios (section 5).

The paper starts with a background on distributed transactions (section 2) and concludes with a discussion of the evaluation (section 6), related work (section 7), further directions for research (section 8) and conclusion (section 9). Evaluation data is available on Zenodo [40].

## 2 Background: Distributed Transactions

Transactions are a mechanism to limit the complexities inherent to concurrent and distributed systems, such as dealing with hardware failure, application crashes, network interruptions, multiple clients writing to same resource, reading of partial updates and data and race conditions [27]. Transactions simplify solving these issues for clients. They group reads and writes together in a logical unit of work, where either all commit, or all abort, even in presence of failures. Transactions can be long running when parties take a long time to respond, for example because of waiting on user input. The safety guarantees for Transactions are ACID [21]: Atomicity, Consistency, Isolation and Durability.

Historically Isolation in ACID guarantees serializability for transactions, meaning operations take effect in a manner equivalent to some serial schedule. However, modern database systems offer a range of isolation properties weaker than serializability [4]. The reason is the trade-off between safety guarantees and performance of the database. Weaker isolation guarantees allow for optimization in performance, especially in a distributed systems setting, where coordination is expensive due to network latency.

This is related to a trade-off in the level of details in the specification of an application. The more that is explicitly known about an application's correctness criteria, the more specific the isolation guarantees can be specialized. In the general case you have to fall back to stronger isolation guarantees. PSAC should simulate the behavior of 2PL/2PC on the object level; since we assume specifications are also on the object level. Strong system-wide guarantees such as serializability are not scrutinized in the current paper, although there is a discussion in section 6.2.

Implementing distributed transactions for distributed objects is the focus of this paper. We use the available semantic knowledge to trade some global isolation guarantees for more local performance, resulting in lower latency and higher throughput.

Distributed Transactions ensure atomicity over multiple application nodes or distributed objects. Two-Phase Locking (2PL) [7, 19, 35] is a concurrency control mechanism and makes sure that serializable isolation is maintained on the application nodes. Two-Phase Commit (2PC) [7, 19, 35] is an atomic commitment protocol that guarantees Atomicity and Durability.

**Concurrency Control and Two-Phase Locking** Consider a bank account object with withdrawal and deposit methods, where a withdrawal should never make an account balance negative. Without concurrency control, it could be the case that two withdrawal actions are simultaneously applied to the account. Both read the same balance and find that individually they do not make the balance negative and are executed, but together they do make the balance negative, violating the invariant. In a serializable situation this is not allowed, since only an outcome state equivalent to a serial execution of both actions would be valid.

Two-Phase Locking (2PL) is a concurrency control mechanism that guarantees serializable isolation (the I in ACID) for a local node or resource. 2PL uses locking to make sure no concurrent changes are made to a resource. It achieves this by using two phases, a growing phase and a shrinking phase in this strict order.

In the withdrawal example, the account resource is locked when the action starts and waits until the first withdrawal action is completed before accepting new actions.

**Atomic Commit and Two-Phase Commit** Two-Phase Commit (2PC) is an atomic commitment protocol that guarantees Atomicity and Durability (from ACID). In itself it does not guarantee Consistency and Isolation. Consistency is achieved by making sure the application invariants are maintained by all operations on the resource. The protocol consists of one Transaction Coordinator and multiple Transaction Participants per transaction. Their internal state is persisted to a durable log, and thus can be recovered in case of failure. The coordinator asks the participants to vote on the transaction. If all participants respond with YES, the coordinator tells them to commit the transactions. If any votes NO, the coordinator tells them to abort. When a participant voted YES, it promises that it will commit when the coordinator requests it, even in case of failures. 2PC is considered blocking, because if the coordinator fails in the specific case when participants have voted, but not yet received a commit decision by the coordinator, the participants are blocked until the coordinator recovers.

**Distributed Transactions**  2PC and 2PL are combined to implement ACID distributed transactions. 2PL's locks are only released when the 2PC transactions are finalized.

2PL locks the resource even though a new incoming transaction might be compatible with the current in-progress transaction, and coordination between the two actions is not actually necessary. This depends on the functional application requirements, which could be less strict than serializability while still maintaining all internal consistency guarantees. The key idea of PSAC is to use available semantic knowledge to determine this, e.g. the outcome of the first withdrawal can never interfere with the acceptance of the second withdrawal when enough run-time balance is available for both. The incoming transaction can be already started, without violating consistency of the balance with respect to its specification. We explore this idea in the next section.

## 3 Path-Sensitive Atomic Commit (PSAC)

In this section we present Path-Sensitive Atomic Commit (PSAC), which exploits statically known preconditions and post-effects to prevent unnecessary locking at run time, and thus increases performance of the overall system in terms of throughput and availability. Intuitively PSAC, like 2PL, is a blocking access protocol between transaction and object, but instead of the opaque "locked" indicator of 2PL, PSAC filters incoming actions which would interact with concurrent actions while letting independent actions through. The strictness of the gate is determined at run time using the possible outcomes of in-progress actions determined by the post-effects, and the preconditions to validate the incoming actions against the outcomes.

Previous work [41] defines independent actions as follows: $IE(e_1, e_2, s) = \forall s' \in State. \, pre(e_1, s) \wedge post(e_1, s, s') \implies \big(pre(e_2, s) \iff pre(e_2, s')\big)$. An action $e_2$ is independent of an in-progress action $e_1$ in run-time state $s$, if and only if its preconditions check result is the same in $s$ and in the post state $s'$, where $e_1$'s effect is applied, e.g. two withdrawals when enough balance is available on the bank account. In order for PSAC to leverage this at run time, the pre- and post-conditions are required to be locally checkable and computable, and totally denote the actions' effects.

PSAC gives the same atomicity and linearizability guarantees as 2PL/2PC, while allowing higher throughput when no local dependency exists. Serializability is not guaranteed, which is discussed in section 6.2. Linearizabilty guarantees an atomic real-time ordering of operations on a single object, as opposed to the global, multiple object-guarantees of serializability. Functional correctness in the local participant is maintained and actions' effects are applied in the original order of arrival.

PSAC combines a variant of 2PL with locks that take the semantics of the actions into account with 2PC. Each resource can have a shared lock when it can be determined that actions are semantically independent. This includes commutative actions. However, even for non-commutative actions PSAC will potentially avoid blocking if actions are independent in the current run-time state, e.g., two withdrawal actions are non-commutative, but will run in parallel by psac if the run-time balance is sufficient for both because neither of them would affect the success or failure of the other one.

psac is faster in accepting actions and increases parallelism when possible, and falls back to the safe 2pl locking approach when not enough information is available. In practice, we limit the number of allowed in-progress actions to be sure that the system can make progress and is not overflowed with accepting new actions on objects. As a consequence, when limiting the maximum number of parallel actions to 1, psac degrades gracefully to standard 2pl/2pc, since new actions are delayed until the single in-progress action's lock clears.

In a scenario with many participants and many requests, but in different transactions (low contention), an application using 2pl/2pc (or psac) is embarrassingly parallel. This means that each of the participants can do their own computations without the need to synchronize with others. These kinds of computations are more easily spread over multiple application nodes.

psac's performance gain over 2pl/2pc becomes evident when multiple actions on the same participant are requested in an overlapping time span. The ability to do parallel processing when application invariants allow it, results in less waiting, and thus more throughput. It also results in processing actions that would otherwise have timed out. This benefit becomes clear at a higher request rate, especially in a higher contention use case, when a few objects are participating in many transactions.

On the other hand, there is also an upper bound to the performance improvement of psac over 2pl/2pc. If the servers running this application are already maxed out on one or more resources, such as cpu, memory or network bandwidth, we expect less improvement, because psac can no longer trade the extra cpu cycles for extra precondition computations and the extra parallel transactions. In the high-contention use-case with a high number of requests, 2pl waits most of the time on locks to clear and many resources are underused. Here lies the biggest performance gain for psac.

### 3.1 PSAC in action

figure 1 and figure 2 visualize the general difference between 2pl/2pc and psac when two actions arrive at the same object in a small time frame. Both figures depict an object sequence diagram. Comment boxes show the internal state of the object, with actions in parenthesis as pending updates. Arrows denote sending and receiving of messages, with withdrawals of €$i$ depicted as '−€$i$'. "apply" and "defer" respectively denote applying of effects and deferring committed effects until later.

Figure 1, on the left, shows the sequence of events when using 2pl/2pc to synchronize. Consider an account object with balance €100 and a precondition check on the withdrawal action that prohibits a negative balance after withdrawal. When withdrawal action $C_1$ (−€30) arrives ①, its preconditions are checked against the current balance. $C_1$ is allowed, the resource is locked and a new 2pc-transaction starts. Even though the account allows the transaction, it is not yet known if the transaction will be committed or aborted by the coordinator, due to processing in other transaction participants. Then another withdrawal action $C_2$ (−€50) arrives ②.

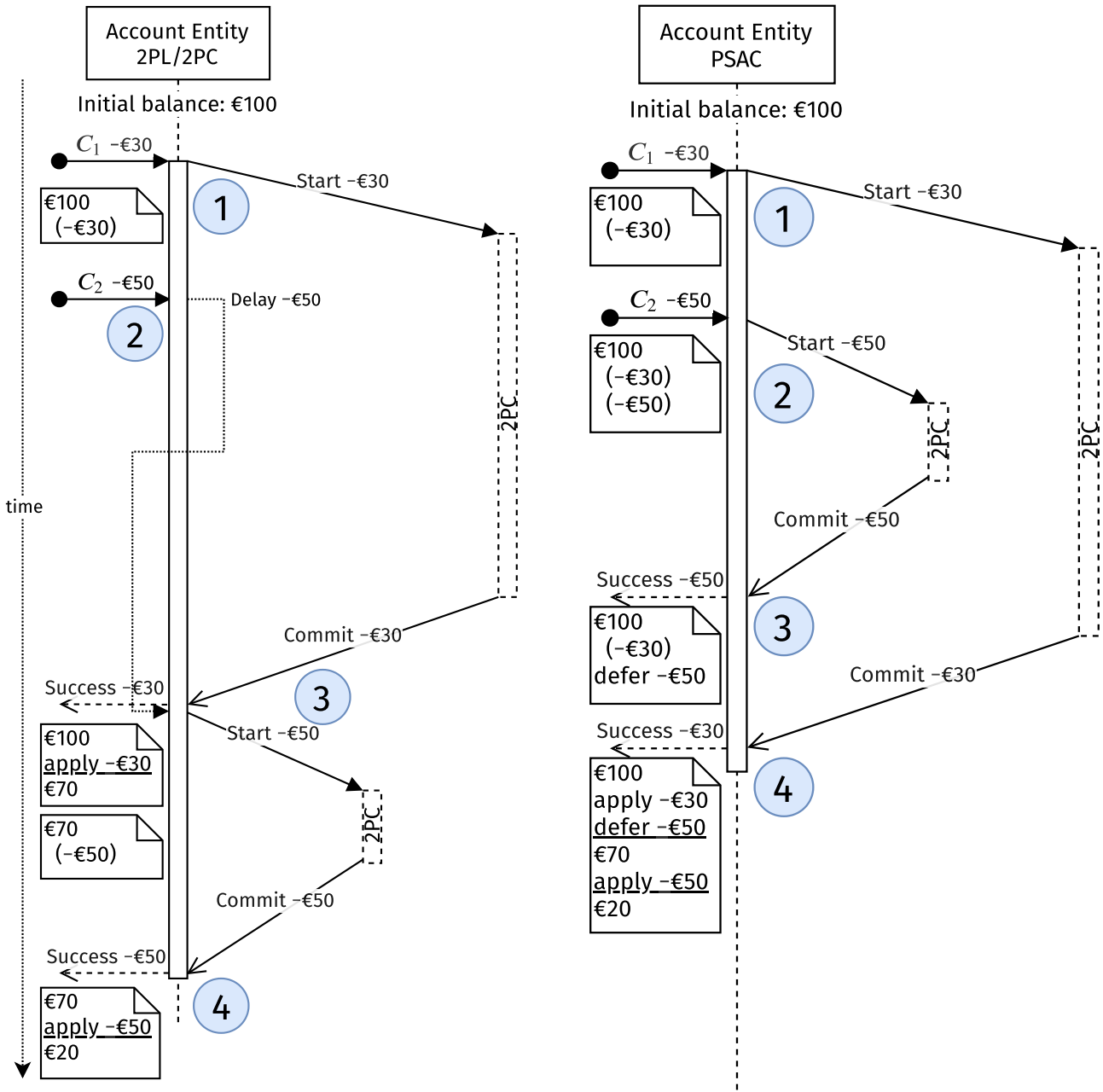

**Figure 1** Vanilla Two-Phase Commit  **Figure 2** Path-Sensitive Atomic Commit

Because the account object is locked, the action is delayed. When $C_1$ commits ③, its effects are applied to the account state, resulting in a new balance of €70 and the object is unlocked. Now, the delayed withdrawal $C_2$ can start, eventually it commits ④ and its effect is applied. This results in the new state of €20. 2PL effectively serializes the two parallel transactions.

The amount of locking performed by 2PL/2PC can be problematic in situations where a lot of transactions happen on a single object. For instance, in the case of ING Bank, when the tax authority pays out benefits to citizens, the bank is required to handle all these transactions within a specific time frame. The tax authority's bank account is highly contended because it is involved with all individual transfers. This would not scale on such an object-oriented message-based distributed system, because each withdrawal will have to wait on the previous to finish.

PSAC improves on this situation by detecting at run time if transactions can be processed in parallel anyway. The same execution scenario is visualized in figure 2, illustrating how PSAC differs from 2PL. We again consider an account object with initial state €100 and a precondition check that prohibits negative balance. Similar to 2PL, when withdrawal action $C_1$ is received ① and no other transactions are in progress, a new 2PC-transaction is started, but contrary to 2PL, the object is not completely blocked. When another withdrawal action $C_2$ arrives ②, it is started because it is independent of whether the earlier action commits or aborts, since there is enough balance to allow the withdrawal to proceed in either case. Therefore, $C_2$ is immediately started. PSAC can detect this independence, based on in-depth knowledge of the functionality of a bank account via the preconditions and post-effects of its actions.

In the example scenario, $C_2$ commits ③ earlier than $C_1$, but its effect is delayed to maintain linearizability of the account. The original requester can be already notified of the successful result (Success −€50), but not yet the new state of the account, since this is dependent on the outcome of $C_1$. $C_2$'s effect is deferred. Now, when $C_1$ commits ④, both effects are applied in order to the account, resulting in a new state of €20.

In situations with non-uniform loads, PSAC delivers on allowing more transactions per time span than 2PL/2PC (and thus higher scalability in terms of throughput). An example with abort is shown in appendix A. We detail the algorithm below and evaluate these claims in section 5.

### 3.2 PSAC Algorithm

listing 1 shows the PSAC algorithm in pseudo-code. The algorithm maintains three lists, inProgress containing transactions that have been started, but have not finished yet; delayed, containing the deferred transactions that have to wait till at least one of the in-progress transactions completes; and finally queued, containing the transactions that are successful, but not yet applied to the state of the object, to maintain the original order of arrival.

On arrival of a command $C_{new}$, its preconditions are checked against all possible outcomes of the transactions that are currently in progress. If it is allowed in all possible states, the action is independent and can start processing. For such transactions it is as if the object is not locked. If there is no possible outcome where the preconditions

▪ **Listing 1** Pseudo-code of a PSAC-enabled object

```
inProgress = []
delayed = []
queued = []

while true:
 if incoming command C_new:
   # See figure 3 for this part of the algorithm.
   S = set of all possible outcome states of C_i ∈
     ↪ inProgress
   if ∀s ∈ S. preconditions of C_new hold:
     inProgress += C_new
     start C_new
   else if ¬∃s ∈ S such that preconditions of C_new
     ↪ hold:
     reply Fail(C_new) to requester of C_new
   else
     delayed += C_new
 else if commit of C_n:
   reply Success(C_n) to requester of C_n
   queued += C_n

 else if abort of C_n:
   reply Fail(C_n) to requester of C_n
   inProgress -= C_n

 C_m = head(inProgress)
 if C_m ∈ queued:
   apply C_m
   inProgress -= C_m
   queued -= C_m
   currentDelayed = delayed
   delayed = []
   for C_i in currentDelayed:
     handle C_i as incoming command
```

of $C_{new}$ hold, the action is immediately rejected with a failure reply. Otherwise, if there is at least one possible state where the preconditions of $C_{new}$ hold, the action is dependent on one of the transactions that are currently in progress, so it is delayed by adding it to delayed. For such a transaction, the semantics of PSAC is equal to 2PL.

Whenever an action commits, it is queued for applying the effects to the object's state. Since all actions are stored in order of arrival in inProgress, it will be applied to the state in the same order. This way non-commutative actions do not violate linearizability. If a transaction aborts, the requester is notified of the failure, and it is removed from the inProgress list. Finally, if the first element of inProgress is in queued, its effects are applied to the state, it is removed from inProgress and queued, and all delayed actions are retried. This results in applying the effects in original arrival order and makes sure delayed actions are retried as soon as possible.

The key idea of the algorithm is the use of the preconditions and actions' effects to construct a tree of all possible outcome states of the set of transactions that are currently in progress. At run time, given the current object state, the set of in-progress actions and the new incoming action, we calculate all possible outcome states of the in-progress actions using the post-effects. This is done by simulating the first in-progress action in the current state, branching into two possible outcomes: one where the in-progress action actually commits and the post-effect is applied, and one where it is aborted and thus not applied. Doing this for all in-progress actions results in a tree with in its leaves the possible outcome states of the object.

figure 3 shows an example of the potential outcome tree $S$ corresponding to the scenario of figure 2 and how it is updated when actions arrive. The leaves represent the potential outcomes. Withdrawal $C_1$ (−€30) arrives ① at a bank account instance using PSAC. The preconditions are valid for $C_1$, and given tentative Abort (−) or Commit (+) by the 2PC transaction, the possible outcome tree branches to two possible outcomes: $S_0$ and $S_{0+1}$, respectively corresponding to a balance of €100 and €70. Withdrawal $C_2$ (−€50) arrives ② and its preconditions are valid in all possible outcomes $S_0$ and $S_{0+1}$, so both possible outcomes branch in similar fashion. Withdrawal $C_3$ (−€60) arrives ③, but its preconditions are *not* valid in all possible states, in particular not in $S_{0+2}$ and $S_{0+1+2}$. $C_3$ is delayed until it is independent from the in-progress actions. In this case $C_3$ is only dependent on $C_2$. The outcome tree is unchanged, since $C_3$ is not accepted for processing yet. When $C_2$ is committed by the 2PC coordinator ④, the possible outcome tree is pruned, because the branches where $C_2$ is aborted are no longer valid, leaving only $S_{0+2}$ and $S_{0+1+2}$. After an in-progress action commits or aborts, in this case $C_2$ ⑤, delayed actions are retried, here $C_3$. Now preconditions fail in all possible outcome states, $C_3$ is independent and thus rejected. When $C_1$ commits ⑥, the possible outcome tree is pruned again and a single state $S_{0+1+2}$ remains. The new state is now calculated by applying the effects in order.

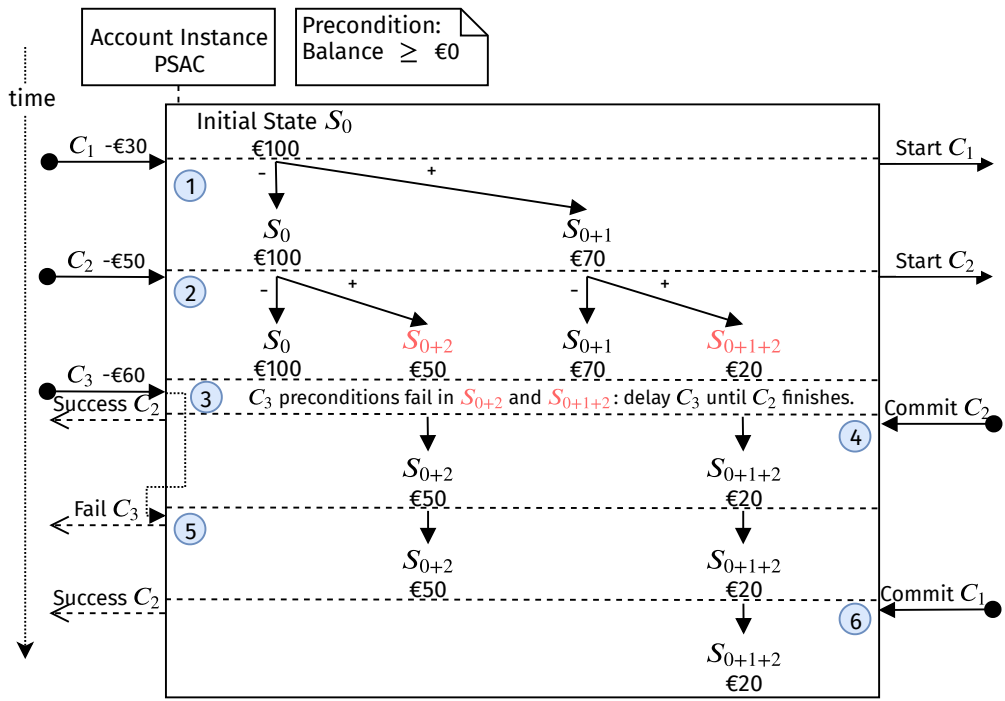

**Figure 3** PSAC example with internal possible outcome tree and decisions on commands

Given all possible outcome states we can check the new incoming action against all outcomes using its precondition. This gives insight if the action conflicts with any in-progress action or combinations thereof. If all or none of possible outcomes satisfy the preconditions, the incoming action is independent and is accepted for processing or immediately rejected.

A difference from 2PL/2PC is that actions that come in later could be accepted for commit earlier. Then the effect of the action is delayed until after the previous actions are committed or aborted, making sure that linearizability of the object is maintained.

## 4  Implementation: Rebel and Akka

To compare PSAC to 2PL/2PC in a realistic environment, we prototyped a small accounting service on top of Akka. For the pre- and post-conditions of transactions we use the Rebel specification language, which aligns with the design requirements of PSAC. Our specific use of Rebel and Akka are not essential to the operation of PSAC but they are part of our evaluation setup for the performance evaluation in section 5.

### 4.1  Rebel: a DSL for Financial Products

Rebel is a domain specific language (DSL) for describing financial products, designed in collaboration with ING Bank, as an experiment to tame the complexity of large financial IT landscapes [43, 44, 45]. Declarative specifications functionally describe financial products, such as current- and savings accounts and financial transfers between them. Rebel specifications are designed to facilitate unambiguous communication with domain experts, support simulation, verification, testing, and execution through code generation. Rebel and proprietary derivatives are used by ING Bank to prototype and understand many different financial products, such as European payments and open data regulations, banking cards and business lending use cases. For example the SEPA specifications consist of 26 Rebel specifications, totaling 964 lines of code.

An example similar to the bank account example used throughout this paper is shown in a Rebel-like specification in listing 2. A specification declares an identity (using the annotation `@identity`), data fields, and describes the life cycle of a product as a state machine with actions and pre- and post-conditions on those actions in predicate logic plus integer constraints.

Listing 2 shows the specification of two classes, `Account` and `MoneyTransfer`. An `Account` is identified by its IBAN bank account number, and has a current balance. The life cycle of an account is as follows: it can be opened, then any number of withdrawals and deposits may occur, and finally it can be closed. Transitions among states are triggered by the actions Open, Withdraw, Deposit, and Close respectively. Each event is guarded by preconditions and describes its effect in terms of post-conditions. For instance, the Withdraw action requires that the withdrawn amount is greater than zero, and that the withdrawal does not produce a negative balance. The effect of withdrawal is then specified as a post-condition on the balance of this account.

The second class `MoneyTransfer` in listing 2 models a transfer of money between two accounts. It can simply be booked via the Book action. The Book action is triggered on two accounts. The effect of booking a money transfer consists of *synchronizing* the Withdraw event on the from account, with the Deposit event on the to account. The `sync` represents an atomic transaction between two or more entities. In other words, an underlying implementation must guarantee that either both Withdraw and Deposit should fail or both should succeed.

The fact that the functional requirements on financial products are formally specified in Rebel separates the "what" from the "how". In other words, decoupling the description of a financial product from its implementation platform allows us to ex-

# Path-Sensitive Atomic Commit: Local Coordination Avoidance for Distributed Transactions

▪ **Listing 2** Rebel specification and state charts of a simple bank account: an `Account` supports events `Open`, `Withdraw`, `Deposit`, and `Close`. A `MoneyTransfer` can be booked by synchronizing `Withdraw` and `Deposit` on two accounts.

```
class Account
  accountNumber: Iban @identity
  balance: Money

  initial init
    on Open(initialDeposit: Money): opened
      pre:  initialDeposit ≥ €0
      post: this.balance ≡ initialDeposit

  opened
    on Withdraw(amount: Money): opened
      pre:  amount > €0, balance - amount ≥ €0
      post: this.balance ≡ balance - amount
    on Deposit(amount: Money): opened
      pre:  amount > €0
      post: this.balance ≡ balance + amount
    on Close(): closed

  final closed
```

```
class MoneyTransfer
  initial init
    on Book(amount: Money, to: Account, from:
         ↪ Account): booked
      sync:
        from.Withdraw(amount)
        to.Deposit(amount)
  final booked
```

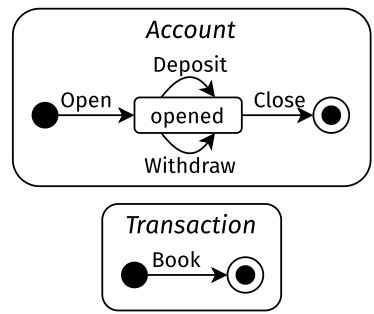

periment with different back-ends for Rebel specifications, by developing different code generators for different platforms or different run-time architectures. Below we show how Rebel classes are mapped to Scala classes that can be executed as actors on the Akka platform. In particular this allows us to experiment with different implementations of the sync construct, such as 2PL/2PC and PSAC.

The consistency of the Rebel classes is fully determined by their life-cycle and pre- and post-conditions and local to the class specification. Isolation guarantees however are undefined for Rebel synchronization [45], although Rebel's simulator and model checker use sequential non-overlapping events, which implies serializability.

## 4.2 Executing Rebel on Akka

**Deployment**  To support fault tolerance and scalability, the execution of Rebel entities is deployed on at least two servers so that customer requests can still be processed when one of the servers breaks down. This means the generated application is a distributed system. One style of implementing a distributed system is by using the actor model [25]. Akka [1] is a well-known toolbox for actor-based systems that runs on the JVM and is widely used to build distributed, message-driven applications. Mapping Rebel objects to Akka actors is a natural fit and provides sufficient low-level controls to vary the implementation of the sync construct. This implementation approach is similar to other reactive architectures such as presented by Debski, Szczepanik, Malawski, Spahr, and Muthig [13].

Each concrete Rebel class instance is run as an actor in isolation and enables distribution of the computation over multiple cluster nodes. Class instance actors are automatically spread over the available cluster nodes to allow for more optimal spread over resources such as RAM and CPU. This enables scaling in and out by moving the actors to other nodes if needed. Each actor runs as an independent object, so it performs work without having to wait on other actors, allowing concurrent work. In theory this means that actor systems scale out linearly, until they have to synchronize. In practice this means that an actor system scales up until too many of its actors are blocked by multiple transactions at the same time.

**Rebel to Akka** Each instance or entity of a Rebel state machine is implemented as an actor. We use the following features of Akka: CLUSTER for cluster management and communication between application nodes; SHARDING for distributing actors over the cluster by sharding on the identity; PERSISTENCE enables event sourcing for durable storage and recovery; and HTTP for HTTP endpoints definitions and connection management. These combined Akka features allow us to spread the Rebel instance actors over a dynamically sized cluster of application nodes. More details on the implementation using Akka are found in appendix C. The back-end for persistence is an append-only event sourcing log, for which we use the distributed and linearly scalable Cassandra database [9].

The runtime guarantees that there is a single actor instance per Rebel class instance and thus guarantees linearizability on instance level, in the sense that operations always see all previous updates. Each operation is persisted to the journal before processing the next, to allow for recoverability and durability in case of failure. The journal data is replicated over three Cassandra nodes. Reads and writes use the built-in QUORUM consistency level of Cassandra to make sure no stale data is read.

An example of the generated Scala code for the `Account` and `MoneyTransfer` example of listing 2 is shown and explained in appendix B.

**Synchronization** We first consider the 2PL/2PC synchronization strategy. Our implementation of 2PC follows the description by Tanenbaum and Van Steen [46] extended with the flattened commit protocol [47] to support nested synchronization in Rebel, where 2PC participants can add more transaction participants. As optimization the transaction manager does not wait on the votes of the other participants and immediately aborts the transaction when one participant aborts. There is a single transaction manager per transaction and one or more transaction participants, respectively implemented by Akka PERSISTENT FSMs named `TransactionManager` and `TransactionParticipant`. They both define a state machine following the definition and also persist their state to the persistence back-end, and thus can be recovered in case of failure.

Both manager and participants have timeouts on their initial states, this means that when no initialization message is received within a given time duration, they will timeout and abort the transaction. This makes sure that the system does not deadlock, although it might result in overhead in creation of transaction actors and messaging when lots of timeouts are triggered.

To make sure no deadlocks happen in other states, timeouts are in place that trigger retries and eventually stop the actor. In the unlikely case that a participant or coordinator is not running, the combination of Akka SHARDING and PERSISTENCE will make sure it is restarted. This also works when some of the cluster nodes shut down, are killed, or become unreachable for whatever reason; in that case other nodes will take over automatically,[2] restore the actors and continue the protocol. The blocking aspect of 2pc, when a transaction manager crashes, is also partly mitigated by message retries and recovering on another application node.

psac is implemented on top of 2pc. Whenever a new action is received by the actor, an action decider function decides if the action can be safely executed concurrently. If the configurable maximum number of parallel transactions per actor is reached, the action is queued. Otherwise, it calculates the possible outcome states by iterating all the possible in-progress action interleavings and checks the preconditions in the calculated states to decide if it can safely start the 2pc transaction for this action. If dependency is detected, the action is also queued. Note that reducing the maximum number of parallel transactions to 1 results in vanilla 2pl/2pc behavior.

## 5 Performance Evaluation

### 5.1 Research Objectives

In this section, we evaluate the performance of psac relative to 2pl/2pc. First, we find out in which scenarios 2pl/2pc is sufficient as a Rebel synchronization back-end and in which scenarios it can no longer maintain sufficient performance. Furthermore, we are interested in determining when psac performs better for the cases where 2pl/2pc is no longer sufficient. In order to look at applications that can scale with business requirements, we focus on scalable and resilient applications that can continue to grow when performance demands keep growing. We study applications that can scale over multiple servers.

The experiments are created to fairly compare psac and 2pl/2pc against each other in the same synthetic scenarios with same load and configuration. We are interested in the scalability of both 2pl/2pc and psac under similar loads. In other words, we are interested in to what extent the throughput increases when more nodes are added to the cluster.

It might seem counter-intuitive that the extra work in psac of calculating the possible outcomes tree and checking the preconditions against all of these states, can result in higher performance compared to 2pl/2pc. For an ideally-scheduled batch based system all extra calculations would worsen performance, since every cpu cycle counts. In this case, the most time in 2pl/2pc is lost by waiting for the unlock. psac's parallel transactions use this otherwise lost time in between for these extra calculations, to determine safe extra parallelization.

We expect that:

**Hypothesis 1.** 2pl/2pc and psac perform similarly in maximum sustainable throughput for actions without synchronization, because objects do not have to wait on each other.

---

[2] The fundamental problem of determining when to fail over, because node failure, slowness and network delay are indistinguishable, is out of scope for this paper.

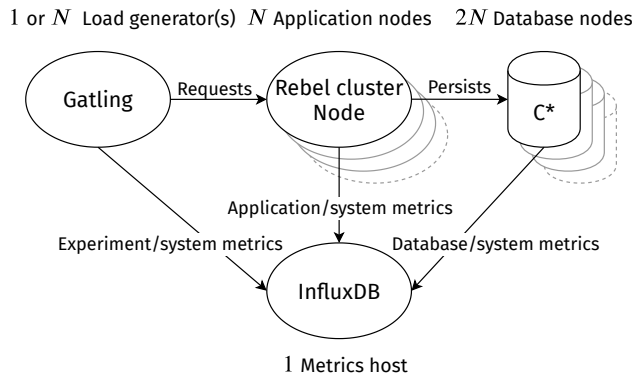

**Figure 4** Experiment setup on $N$ nodes. A single or more load generators do HTTP requests randomly over $N$ Rebel application nodes, backed by $2N$ Cassandra database nodes. Relevant experiment and system metrics are reported to InfluxDB for later analysis.

**Hypothesis 2.** 2PL/2PC and PSAC perform similarly in maximum sustainable throughput for actions with low contention synchronization, because synchronization is evenly spread over the objects.

**Hypothesis 3.** PSAC performs better than 2PL/2PC in maximum sustainable throughput for actions with high-contention synchronization, because 2PL/2PC has to block for in-progress actions, where PSAC allows multiple parallel transactions.

Before it can be determined if PSAC is generally useful, first we need to find out whether PSAC pays off in high-contention scenarios. Since PSAC is a new algorithm running a complex technological context the answers to these hypotheses are not trivial: first the expected gain may not be significant compared to other relevant factors and second the cost of the additional overhead for every transaction may outweigh the benefits. So, these experiments are designed to first isolate the effect of PSAC as compared to 2PL/2PC, and then to try and invalidate the above hypotheses. If the experiments can not invalidate our claims, then we gain confidence in the relevance of the new algorithm.

To be sure that we are not running into the limits of (configuring) the infrastructure, but really into limits of the synchronization implementation, we investigate first how far we can take the Akka infrastructure without any logic or synchronization.

**Hypothesis 0.** The actor system infrastructure enables horizontal scalability, which means that adding more compute nodes increases throughput.

## 5.2 Deployment Setup

In order to scale to multiple nodes, our experiment setup runs on Amazon ECS (Elastic Container Service) using Docker images for the Database (Cassandra), the Application, Metrics (InfluxDB), and the Load generator (Gatling [17]). Figure 4 shows an overview of the setup. The Cassandra version is 3.11.2 on OpenJDK 64-Bit Server VM/1.8.0_171. The application runs on Akka version 2.5.13, Oracle Java 1.8.0_172-b11, with tuned garbage collector G1 with MaxGCPauseMillis=100.

In order to prevent CPU or memory starvation/contention between the application and the load generator tool, we deploy each of the application components on a different virtual host on Amazon Web Services (AWS). We use EC2 instance type M4.XLARGE[3] for all VMs, which are located in the Frankfurt region in a single data center and availability zone.

Each of these containers is deployed on its own container instance (host), with the exception of Metrics and Load generator, which share a host. Metrics being sent asynchronously over UDP, to ensure minimal interference with application performance. CPU and other system metrics are monitored to prevent this.

For realism of the experiments we use the production-ready persistent journal implementation Cassandra as an append-only log for the persistent actors, so limited synchronization is done on the database level, although it gives realistic overhead. We over-provision the database to make sure it is not a bottleneck.

Our tooling supports running the performance load from multiple nodes. Experimentally we discovered that setting up the correct experiment for high load is not trivial: such as the correct number of file descriptors for connections; garbage collection tuning; library versions with bugs; careful load generation to capture the sustainable throughput; ratio of application, load and database nodes; collection of metrics for all components; and validating correct deployment before running the experiment. We collect system metrics for all machines in order to monitor overload of any specific part. The low-overhead JDK Flight Recorder profiling is also enabled for after-the-fact bottleneck analysis of our application nodes. The experiment metrics results and profiling files are available at Zenodo [40].

When load testing applications, the crafting of the load is very important, and not trivial. A distinction often used is closed versus open systems [39]. Closed systems have a fixed number of users, each doing requests to the service, one after another, limiting the total number of TCP connections. Open systems have a stream of users requesting at a certain rate, meaning there is no such maximum of concurrent requests as in a closed system. Typically closed systems are used for batch systems and open systems for online usage.

For all experiments presented in this paper, we employ a closed system workload approach. Finding the maximum throughput using an open-world workload quickly results in an overloaded application, both for 2PL/2PC and PSAC, which obscures the differences between them. In an enterprise setting, such as a bank, a (hardware) load balancer translates the open workload behavior to a more closed world behavior by limiting the number of network connections and reusing them.

Each request from the load generator to the application will spawn a 2PC coordinator actor for the request a 2PC participant actor for each synchronization participant. For the bank transfer experiments, this means that for each request, a new Rebel MoneyTransfer entity actor is started, one 2PC coordinator actor, and three 2PC participants (for the money transfer and the two accounts). So the number of actors created

---

[3] M4.XLARGE: 4 vCPU, 16 GiB Memory, EBS-based SSD storage, 750 Mbit/s network bandwidth.

■ **Table 1** Baseline experiment fit to Amdahl's law and asymptote

| experiment  | $\lambda$ (tps) | $\sigma$  | $a_{inf} = \lambda\sigma^{-1}$ (tps) |
|-------------|-----------------|-----------|--------------------------------------|
| Bare        | 16 751          | 0.002 923 3 | 5 729 998                          |
| Simple      | 10 372          | 0.000 877 3 | 11 822 028                         |
| Sharding    | 6303            | 0.004 728 5 | 1 332 920                          |
| Persistence | 1928            | 0.008 159 7 | 236 281                            |

is roughly five times the number of requests. For our experiment scenarios all actors are equally spread over all the Akka cluster nodes.

### 5.3 Baseline Experiments: Akka Scalability

To make sure that Akka or our setup does not influence the result of evaluating the performance of PSAC, we run four experiments on top of plain Akka to establish horizontal scalability. The goal of this experiment is to isolate (environment) noise and reduce confounding factors. The baseline experiments use a setup as similar as possible to the more involved experiments discussed later. We run multiple variants that increase in complexity, building up to all the features used by the Rebel implementation, and measure the maximum sustainable throughput (requests/transactions per second) per each increment of application complexity.

The following experiments were run:

1. Bare – HTTP: responses are immediately given by the HTTP layer.
2. Simple – HTTP + Actors: each request creates an actor which sends the response.
3. Sharding – HTTP + Sharded Actors: actors are equally spread over the cluster and send the response
4. Persistence – HTTP + Sharded Persistent Actors: actors are spread equally over the cluster and wait for a successful write to the persistence layer (Cassandra) before responding to the request.

The application responds with a JSON message when the work is described is done. A request is successful when a 200 HTTP status code is received.

Figure 5 shows the throughput results of the experiments. Data points are throughput in terms of successful responses per second during the stable load of the experiment, after warm-up and ramp-up of users. For warm- and ramp-up we increase the number of simulated users over time, to give the application some time to get up to speed. The plot also shows a fit to Amdahl's [2] law using a non-linear least squares regression analysis. For intuitive comparison we include the upper bound of linear scalability line for each of the experiments. Amdahl's law is defined as: $X(N) = \frac{\lambda N}{1+\sigma(N-1)}$, where $X(N)$ is the throughput when $N$ nodes are used. Linear scalability means that the contention $\sigma$ is 0 and the throughput grows with $\lambda$, which denotes the throughput of the single application node. The fitted values for $\lambda$ and $\sigma$ are shown in table 1.

All variants have very different performance per node. This is expected, by the increasingly complex actions performed. Increasingly complex variants have increasing $\sigma$, which can be explained by increased synchronization between the Akka application nodes. All experiments show horizontal scalability up until an expected peak throughput on Amdahl's law asymptote ($a_{inf} = \lambda \sigma^{-1}$), which is the theoretical maximum throughput which can not be further improved by adding more nodes.

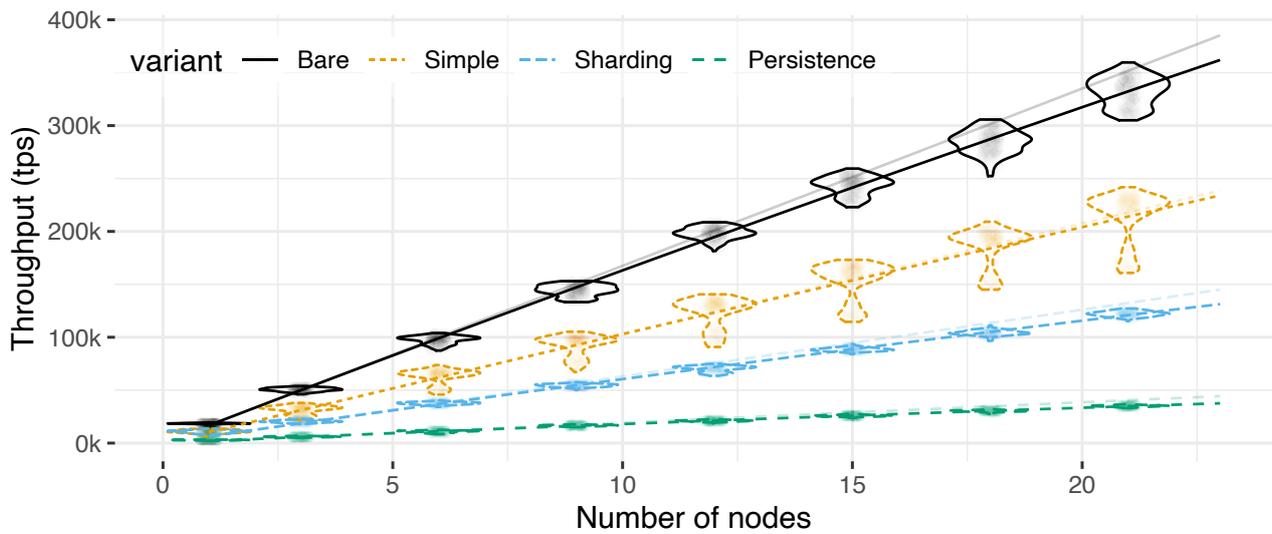

**Figure 5** Throughput $X(N)$ (violins), Amdahl fit (colored line) and linear scalability upper bound (transparent line) of baseline experiments

The results show that our implementation using Akka exhibits horizontal scalability and corroborates hypothesis 0.

### 5.4 Synchronization Experiments: PSAC vs 2PL/2PC

To compare the performance of PSAC and 2PL/2PC, we run three experiments with different synchronization characteristics, linked to the relevant hypothesis:

1. NoSync – OpenAccount: A Rebel operation without sync. (Hypothesis 1)
2. Sync – Book: A Rebel operation with sync, synchronizing with Withdraw and Deposit on two accounts. (Hypothesis 2)
3. Sync1000 – Book on a limited number of accounts, to increase the contention. (Hypothesis 3)

These different scenarios enable us to see if and when PSAC improves over 2PL/2PC, especially in the Sync1000 high-contention experiment. On the one hand NoSync and Sync show where PSAC performs similarly 2PL/2PC. On the other hand Sync1000 shows the high-contention scenario where PSAC improves over 2PL/2PC.

All three experiments use a closed system approach [39], where we limit the number of concurrent total users. This ensures that the application is not overloaded by too many requests, causing high failure rates. Each experiment is run consecutively for increasing node count $N$, with $N$ load generator nodes (except Sync1000) to grow the load proportionally. Sync1000 runs a single load generator which increases the load in incremental steps in order to determine the maximum throughput until the application overloads.

The high-contention scenario Sync1000 is designed to be as close as possible to a realistic industry setting, where high-contention objects become a bottleneck. This is similar to the NewOrder benchmark of the well-known TPC-C [37] online transaction processing benchmark suite, where a high-contention object is responsible for handing out order IDs.

In all experiments we compare 2PL/2PC's and PSAC's throughput ($X(N)$) for a varying number of application nodes $N$.

**NoSync** The NoSync experiment is the Open Account scenario which does not contain a Rebel sync. It corresponds to hypothesis hypothesis 1, which states that 2PL/2PC and PSAC should have similar throughput when there is no synchronization for the actions involved. The results are plotted in figure 6a. We observe that the throughput of the two variants is similar, as expected and thus corroborates hypothesis 1. The throughput is only limited by the CPU-usage on the nodes and the creation of records in the data store. The metrics data shows that the application CPU usage drops to around 80 % and the data store CPU usage is almost 100 %.

**Sync** The Sync experiment contains a sync in the Book action and corresponds to hypothesis 2, which states that we expect that PSAC and 2PL/2PC also have similar throughput in this low-contention scenario. The results are shown in figure 6b. Here we also see the same performance for both 2PL/2PC and PSAC, corroborating hypothesis 2. This can be explained by the experiment setup: The Book actions are done between two accounts uniformly picked from 100.000 accounts initialized before the experiment. With a maximum throughput of roughly 1500 and uniformly spread bookings the probability of overlapping transactions on a single account is low.

The absolute throughput numbers are lower than NoSync, however, which is explained by the fact that Book has to do more work, since it involves three instances: one `MoneyTransfer` and two `Account`s.

**Sync1000** Finally, Sync1000 introduces artificial contention by reducing the number of accounts to 1000, corresponding with hypothesis 3. Hypothesis 3 states in high-contention scenarios that PSAC is expected to have higher throughput than 2PL/2PC, because it is able to avoid blocking where 2PL/2PC can not. This results in a difference between 2PL/2PC and PSAC, as seen in figure 6c. Since this is the most interesting case we have run the experiment for higher node counts, and include a fit on Amdahl's law, shown in figure 7. Figure 6d contains the fitted parameters. The results show that PSAC consistently achieves higher throughput than 2PL/2PC.

The metrics show that both application and data store CPU usage starts dropping for node counts $> 9$. This can be explained by contention: busy entities are at their maximum throughput for 2PL/2PC transactions. In the case of PSAC this also happens, because the number of parallel transactions is limited by configuration at 8. Nevertheless, PSAC consistently achieves higher throughput.

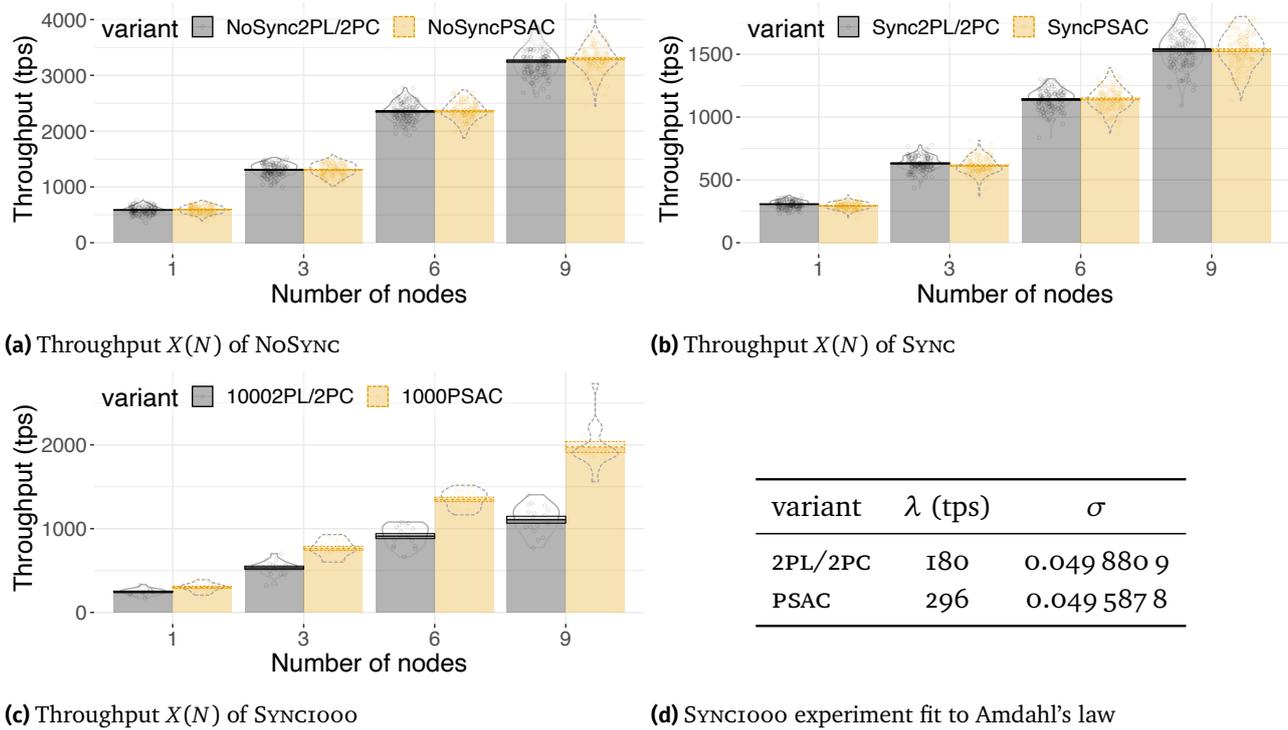

**(a)** Throughput $X(N)$ of NoSync

**(b)** Throughput $X(N)$ of Sync

**(c)** Throughput $X(N)$ of Sync1000

**(d)** Sync1000 experiment fit to Amdahl's law

| variant | $\lambda$ (tps) | $\sigma$ |
| --- | --- | --- |
| 2PL/2PC | 180 | 0.049 880 9 |
| PSAC | 296 | 0.049 587 8 |

■ **Figure 6** Throughput $X(N)$ against number of nodes $N$. This pirate plot is a combination of violin plot, box plot and bar chart. Line is the median, points are the data points. This gives a complete overview of the data (data points and distribution in violin plot) and an aggregated view.

The graphs in figure 8 display the latency percentiles against increasing throughput. Since the Y-axis of the different graphs is the same, we can see that the latencies for all node sizes are similar, but the throughput grows larger when node size increases. This also shows clearly that PSAC reaches higher throughput levels and that PSAC is on par or better latency-wise with 2PL/2PC up to at least the breaking point of 2PL/2PC, which is explained by the improved parallelism on PSAC.

## 6 Discussion

### 6.1 Threats to Validity

We distinguish between construct validity, internal and external threats to validity. Construct validity discusses if the test measures what it claims to measure. Internal threats are concerned with problems of configuration and bugs in the implementation. External threats are about the generalization of the results.

**Contruct validity** Regarding construct validity we have mitigated this risk by first doing a infrastructure and a NoSync experiment (hypothesis 0), in order to make sure that we actually measure the intended construct: comparing PSAC against 2PL/2PC. How sure can we be that in our situation with a noisy cloud environment, the results are significant compared to coincidental variation? Our experiments compare between variants under the same benchmark and implementation to be sure we are correctly comparing the relevant synchronization implementation parts of the setup. The baseline experiment (hypothesis 0) makes sure that the setup and environment are correctly configured, and provides bounds in throughput and latency in which the results of the actual experiments are to be interpreted. The Sync1000 experiment is set up in such a way that if PSAC did not significant improve performance, this would be visible in the results. The other experiments (NoSync, Sync1000) are its baseline to show PSAC's and 2PL/2PC's variance is limited in other (low-contention) situations. This shows that PSAC's performance improvement in the high-contention scenario is not due to noise or external factors.

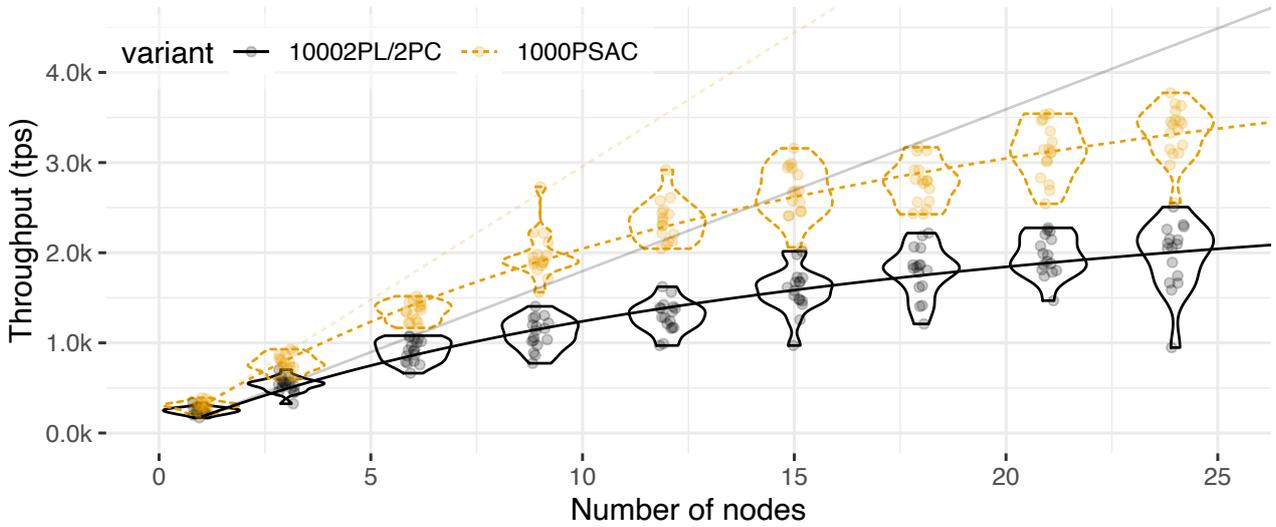

**Figure 7** Plot of Amdahl fit and corresponding linear scalability upper bound (transparent) for Sync1000 on higher node counts

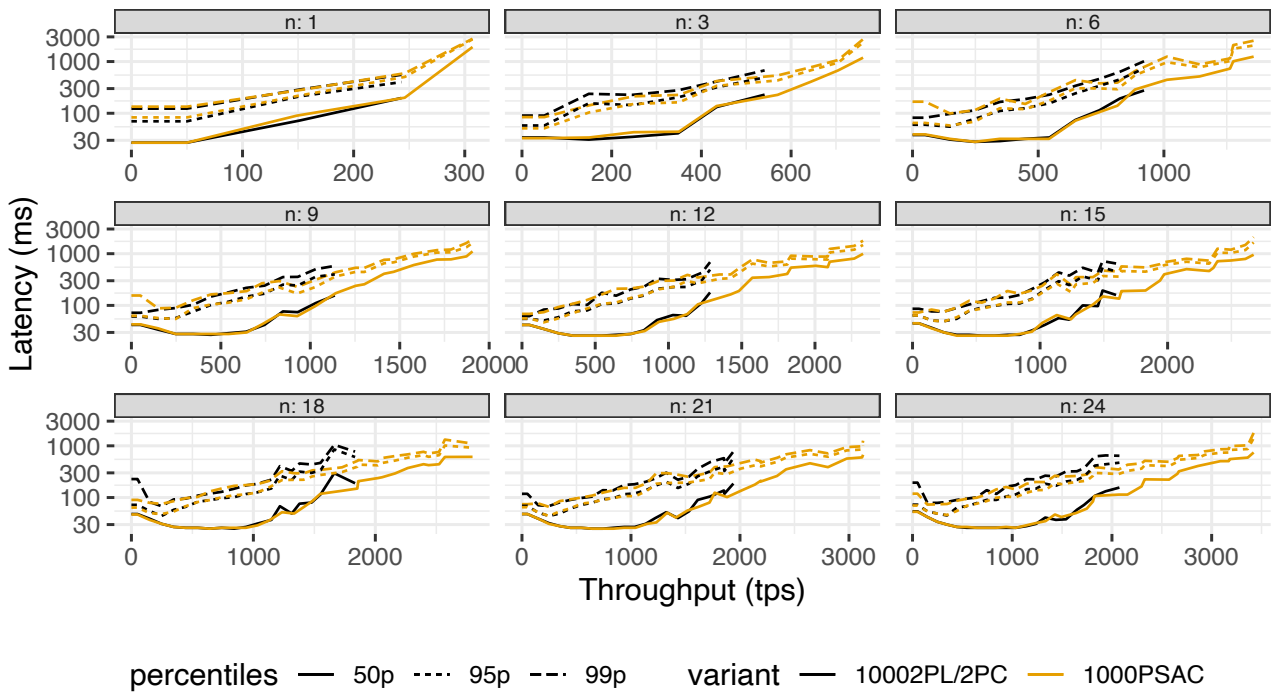

**Figure 8** Latency percentiles (logarithmic scale) of Sync1000 grouped by number of nodes (n). Y-axis shared for latency comparison, lower is better. X-axis: higher is better.

**Internal threats to validity**  To make sure there are no differences in configuration and deployment of our experiments, we designed and implemented an experiment runner to automatically run the different scenarios required for each experiment on the available AWS VMs. The experiments are defined using declarative configuration, to make them reproducible and without configuration mistakes. For each experiment each node size is run separately on AWS. The use of Docker images and automated tooling makes sure that the configuration and artifacts for each of the experiments are the same, except for the specific differences that we want to compare.

Another threat is the Amazon virtual machine environment: this can be a noisy environment, which influences our experiments. Nodes are run on possibly shared hosts, which may impact performance depending on noisy neighbors, differences in hardware, or even time of day. Warm up time is frequently the bottleneck in data parallel distributed systems on the JVM [30], so this factor may not be eliminated by our experiments. Experiments may also not have been run long enough to obtain reliable results. We mitigated this threat partially by (a) designing our experiments to compare between variants under the same conditions and (b) running the experiments on many different occasions and manually validating that the results are similar to previous runs. There is a threat that our findings do not generalise to a broader range of scenarios.

Another possible influence on the performance results is the persistence layer. In order to make sure the persistence layer is not the bottleneck, we should monitor metrics of the database nodes, such as CPU, IO and memory usage. If none of them continuously peak, we assume this is not a bottleneck. However, during the execution of some of the experiments the persistence layer has not been monitored consistently.

**External threats to validity**  For PSAC to be correct and consistent, the defined pre- and post-conditions have to be precise and fully define the checks and effects of an implementation. In practice PSAC's implementation uses the same non-side-effecting code to calculate the possible outcomes as for the actual state changes. When PSAC is used as part of another implementation, care has to be taken.

The load might be too hard on the system, resulting in higher throughput but worse response times than we want. This could obscure comparison and generalization. For instance, the SYNC1000 experiment for PSAC showed overall higher throughput, but also increasing latencies. We expect that tuning of the load reduces the pressure on the application and will result in improved latencies to 2PL/2PC but at higher throughput.

The experiments reported on in this section are still relatively isolated. In order to claim generalized applicability, further work is needed to obtain results in different settings, and different kinds of loads. Related work [41] studies statically independent events, which is a subset of the independent actions discussed in this paper. They show that at least 60 % of event pairs in state machine models from industry can benefit from independent actions. To show PSAC's performance gains in real-life scenarios, orthogonal research is needed to show that these independent actions occur in high-contention scenarios. For instance, it would be interesting to see how PSAC performs on some well-known benchmarks, such as TPC-C [37], the twitter-like Retwis Workload [29], YCSB [10], the SmallBank benchmark [8] and the OLTPBench benchmark runner [14]. Modeling TTPC-C's NEWORDER is non-trivial in Rebel, because of a mismatch with SQL transactions, which can contain multiple queries and updates based on each other, where a Rebel event is non-interactive.

A geo-located setup, furthermore, would make the experiments more realistic, because round trip times to application nodes and database nodes are relatively large. We expect contention to be more of a problem there, because the latency of individual transactions (and thus the amount of locking) goes up. PSAC could be extended to employ techniques similar to Explicit Consistency [6] to allow parallel multi-regional actions without immediate communication.

### 6.2 Limitations

PSAC results in performance gains when actions are independent and there is much contention. So in practice the benefit depends on the use case, because it might not be a high-contention scenario. PSAC's benefit is most clear in the situation where (a) objects are involved in many (long running) synchronized actions from different other objects, making that single object a bottleneck for the others and (b) when all actions being used for synchronization are independent. This situation results in a scalability limiting factor where PSAC improves throughput and latency performance over 2PL/2PC. In the case with many objects which do not interact via synchronization or the request volume is low, PSAC's performance gain is limited, although never worse than 2PL/2PC in the same situation, as shown by the NOSYNC and SYNC experiments.

A current limitation of PSAC is that it does not offer fairness for dependent actions. PSAC accepts new independent actions when there are also dependent actions in the wait queue. In a pathological scenario this results in a new in-progress action that keeps the queued action dependent, and thus will potentially never be removed from the queue. A potential solution is to consider the dependency of the queued actions on the incoming action, when determining independence, so that queued actions are never requeued indefinitely. Another, simpler but less fair, solution is to make sure only a limited number of independent actions can go before the dependent action.

**Path-Sensitive Atomic Commit: Local Coordination Avoidance for Distributed Transactions**

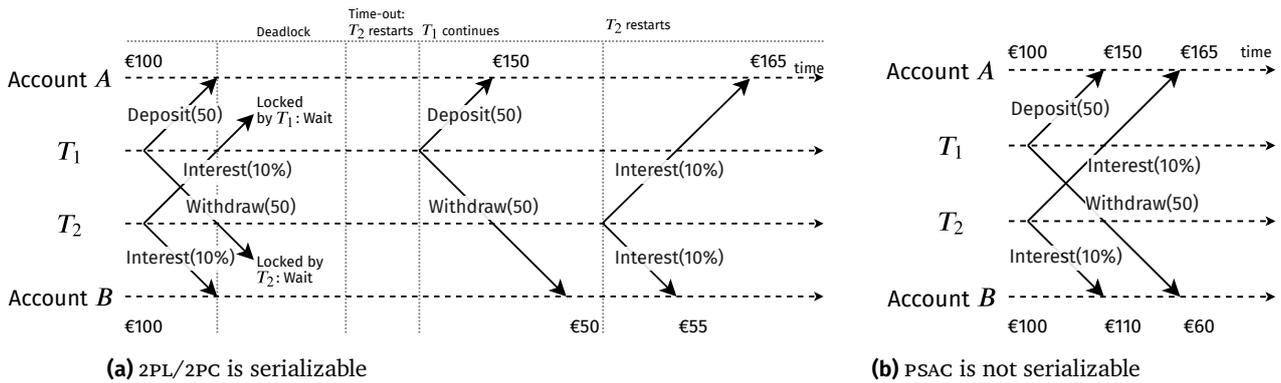

**Figure 9** Example with two events, to show difference in isolation between 2PL/2PC and PSAC

**PSAC is not Serializable** PSAC does not guarantee serializability, while 2PL/2PC does. Consider the following situation, as shown in in figure 9: Two distributed transactions $T_1$ and $T_2$ are concurrently started. $T_1$ consists of two actions, Deposit(50) and Withdraw(50) on respectively Account $A$ and Account $B$, both with a balance of €100. $T_2$ consists of two actions Interest(10 %) on both Account $A$ and Account $B$. $T_1$ first arrives at Account $A$ and $T_2$ arrives first at Account $B$.

For strict 2PL (see figure 9a) this means that $A$ is locked by $T_1$ and $B$ is locked by $T_2$. Now both transactions are waiting on the other Account to acquire locks. This deadlock situation is solved by a deadlock mechanism, such as timeouts: one of the two transactions times out, its lock is released and the other makes progress.

In this situation PSAC will allow both transactions to take a shared lock (see figure 9b), since for each transaction the already in-progress event's outcome does not influence the validity of the precondition of the other: Interest(10 %) is valid on Account $A$, regardless of the commit or abort of Deposit(50). The same holds vice versa for Account $B$. Both transactions commit and have their effects applied. For Account $A$ this results in applying the effects in order or arrival, first Deposit(50), then Interest(10 %): $(100 + 50) * 1.10 = 165$. For Account $B$ first Interest(10 %), then Withdraw(50): $(100 * 1.10) - 50 = 60$. Both accounts are in a valid state according to their specification, but notice that the transactions are applied in different order for the accounts. A first applies $T_1$, then $T_2$. B first applies $T_2$, then $T_1$.

For a valid serializable schedule for the whole system, in this case the two entities, the resulting state should be equivalent to an outcome state of a sequential execution of all transactions. Serializability requires one of two possible histories: $\langle T_1, T_2 \rangle$ or $\langle T_2, T_1 \rangle$. The results of these histories are respectively:

$\{A: 100, B: 100\} \to^{T_1} \{A: 150, B: 50\} \to^{T_2} \{A: 165, B: 55\}$ and
$\{A: 100, B: 100\} \to^{T_2} \{A: 110, B: 110\} \to^{T_1} \{A: 160, B: 60\}$

The outcome for PSAC in this situation, $\{A: 165, B: 60\}$, which is not one of the valid serializable configurations. Ergo, this counter example shows that PSAC is not serializable. Determining the isolation guarantees of PSAC more precisely is part of future work.

### 6.3 Evaluation

We have seen in the previous section that PSAC outperforms 2PL/2PC in throughput and its request latency is on par or better. This is due to less locking by PSAC, which is isolated by the NoSync experiment, where both PSAC and 2PL/2PC perform similarly when no transaction have to wait due to locking. However, PSAC does not give the same serializable isolation guarantee as 2PL/2PC. In our experiment scenarios with withdraw and deposit events this does not lead to different results or outcome states as in a serializable schedule, because these event's effects are commutative and result in serializable histories with PSAC.

## 7 Related work

**Distributed Transactions in Actor Systems** Orleans [34] is an actor based distributed application framework that implements transactions [15] in a similar way to 2PL/2PC, but with a central Transaction Manager, which decides if transactions are incompatible. To support high throughput the distributed object releases the 2PL lock when it prepares successfully, and already applies the new state. If a transaction triggers an abort, all the actions on top are also aborted (cascading abort). This solution enables high throughput, but it drops when aborts happen regularly on congested instances.

Reactors [38] is a distributed computing framework defined on reactors: actors reacting to events. It uses Reactor transactions with nested sub-transactions, but is not yet tested in a cluster of nodes. Their current implementation also uses 2PC in the transaction manager.

**Coordination** More recent work in distributed systems is investigating requirements to keep a program functionally correct, instead of focusing on data consistency (or memory consistency) where registers with single data items are always in a consistent state, which is what 2PL ensures. The CALM paper [23] hints at creating programs that are monotonic by construction, by using languages that help monotonic specification. PSAC makes sure objects only increases monotonically on the life-cycle lattice of an entity as defined by its specification. Parallel events are only allowed when the functional application properties (pre-/post-conditions) allow this. This makes sure that entities are monotonic by construction, w.r.t. their specification. This is a step towards CALM in the sense that it allows designers to write specifications with coordination, which in the end are run without local coordination by PSAC.

ROCOCO [31] reorders transactions at run time, whenever possible, instead of aborting. It uses offline detection, but only works on stored procedures. Coordination Avoidance [3, 5] focuses on lock-free algorithms in a geo-replicated setting. It makes sure that transactions do not conflict, and allows them on multiple geo-located data centers without coordination. They are eventually merged in an asynchronous fashion. Bailis [3] states: "Invariant Confluence captures a simple, informal rule: coordination can only be avoided if all local commit decisions are globally valid." PSAC focuses on local avoidance of coordination of transactions on objects and it is yet to be seen how well it works in a geo-replicated setting.

PSAC is based on detecting independence of actions at run time. A compatible approach [41] to avoid blocking is to use static analysis of pre- and post-conditions to determine whether certain types of actions are always independent of other types of actions for all possible run-time states and action field values. Actions which never influence the outcome of later actions, such as depositing money in the running example, can always be safely started in parallel, without checking all possible outcomes of in-progress actions.

Using commutative operations to reduce coordination is a productive area [3, 5, 6, 18, 23, 32, 36, 48]. Commutative operations always result in the same outcome state, even when the operations are reordered. These works prevent coordination by relying on reordering and commutativity of operations in order to allow parallel operations in mainly geo-distributed data center environments.

**Other related work**   The Escrow Method [33, 47] is a way to handle high-contention records for long running transactions. Balegas, Duarte, Ferreira, Rodrigues, Preguiça, Najafzadeh, and Shapiro [6] discuss Escrow reservations with numeric fields divided over multiple (geo-located) nodes. Each node locally decides up to a maximum amount, and communicates with the rest when it needs more. In the banking example this is analogous to splitting the balance of an account in parts and allow nodes to locally mutate that part without synchronization. Although PSAC is not optimized for geo-separation, since an object is not divisible in multiple parts, it is not limited to numeric fields.

PSAC is related to Predicate locks [16, 20, 26] and Precision locks [20, 26], but differs in the sense that the latter operate at the level of tuples. PSAC supports more granular locking because two independent actions can change the same field or tuple.

Phase Reconciliation [32] is a run-time technique that splits high-contention objects over multiple CPU cores. It allows specific commutative operations of a single type to be processed locally on the core in parallel and after a configurable window the results are reconciled again. PSAC operations also cannot return values, however PSAC does not require commutative operations or all operations to be of a single type.

Flat Combining [24] is a technique to speed up concurrent access to data structures. The first thread to get the lock on a shared data structure, processes the operations of concurrent operations in a batch and informs the requester threads of their respective results, resulting in improved throughput. PSAC focuses on distributed transactions, where the actual transition is determined externally from the object by a transaction manager and not on applying operations sequentially as fast as possible.

## 8   Further Directions

In this paper, we have presented PSAC informally. Further research is needed to obtain precise results about the isolation guarantees that PSAC offers. A potential direction could be to formally verify the correctness of PSAC, for instance, using TLA+ [28], or state-based formalization [11].

Further, the implementation of PSAC could be improved by applying well-known optimizations of 2PC. For instance, using half round trip time locks [3] the set of participants is forwarded by the previous participant to the next, in a linked list-like fashion. This results in half the round trip time for acquiring the locks compared to the approach where locks are acquired one-by-one by the transaction coordinator.

Additional optimizations are possible in the representation of the outcome tree. For instance, outcomes could be grouped by abstractions, such as minimum or maximum values, sets of outcomes, or predicates deduced from pre- and post-conditions. This reduces the size of the tree, and thus faster precondition checking.

PSAC can be further improved by reordering of actions, however this requires commutative operations. At run time it can be checked if an incoming action is commutative with all in-progress actions, and safely reordered. However, the pre- and post-conditions should be explicit about time sensitive or otherwise important functional action ordering.

The depth of the possible outcomes tree is limited by configuration, because it grows exponentially in the number of in-progress actions. Benchmarks, not shown in this paper, show that when it grows too big for the bank transfer use case, actions start timing out. This performance impact of varying this depth greatly depends on how computationally expensive actions are and how much contention there is, but also how many other resources are running and their contention. More in-progress actions, result in more running actors and extra calculations of the tree. It is future work to find an approach to tune this tree depth.

In order to evaluate the boundaries of applicability of PSAC, an extra experiment that tries to maximize the overhead of PSAC can be created. If computing preconditions or post states is expensive, the extra calculation overhead could result in worse performance than the sequential 2PL/2PC approach.

For online user experience keeping tail latencies low is important. We can apply the techniques presented in The Tail at Scale [12] to make PSAC more latency tail-tolerant. This requires sending multiple omnipotent requests to different application nodes and effectively increasing replication factors of the entities. The current design does not fit this yet, since the runtime makes sure only one instance of each entity is alive in the cluster. PSAC can be extended, however, to support read-only versions, inferring actions that are commutative to be applied on different nodes in arbitrary order order.

## 9 Conclusion

Large organizations such as banks require enterprise software with ever higher demands on consistency and availability, while at the same time controlling the complexity of large application landscapes. In this paper, we have introduced path-sensitive atomic commit (PSAC), a novel concurrency mechanism that exploits domain knowledge from high-level specifications that describe the functionality of distributed objects or actors. PSAC avoids locking participants in a transaction by detecting whether requests sent to objects can be handled concurrently. Whether the effects of two or more requests are independent is established by analyzing the applicability and effects of message requests at run time.

psac has been implemented in the actor-based back-end of Rebel, a state machine-based DSL for describing business objects and their life cycle. Rebel specifications are mapped to actors running on top of the Akka framework. Using different code generators this allowed us to explicitly compare standard 2pl to psac as locking strategies for when objects need to synchronize in transactions. We conducted an empirical evaluation on an industry-inspired case of psac compared to an implementation based on standard two-phase commit with strict two-phase locking (2pl/2pc). We designed multiple experiments to show specifically where psac and 2pl/2pc perform similar and where psac outperforms 2pl/2pc.

Our results show that in low contention scenarios with and without synchronization the throughput is similar, because no actions can be parallelized. However, psac performs up to 1.8 times better than 2pl/2pc in terms of median throughput in high-contention scenarios. This is especially relevant, for instance, when a bank has to execute a large number of transactions on a single bank account. Latency-wise psac is on par or better than 2pl/2pc. Furthermore, psac scales as well as 2pl/2pc, and under specific non-uniform loads even better.

## A  Example 2PL/2PC and PSAC diagrams with ABORT

Figures 10 and 11 show the same example situation as used in section 3.1, but in this case the first action 2PC transaction aborts. We see at figure 10.③ and figure 11.④ that action −€30 is aborted by the 2PC coordinator. With PSAC −€50 still starts, since both outcomes (commit and abort) where taken into account.

## B  Actor class definition

The library using Akka expects the Rebel specifications to implement a Scala trait RebelSpec. A simplified version of the Scala code generated from the Account example of listing 2 is shown in listing 3.

The algebraic data types AccountState and AccountCommand model respectively the Rebel state machine states and actions . The methods initialState, allStates, and finalStates encode metadata of the life cycle of an entity. The method nextState encodes how transitions are performed and via which events. The preconditions and actions' effects required for PSAC are known from the Rebel specification and generated into the actor code. checkPre checks the preconditions for each incoming action. As a result, the method apply calculates the new state of the account given the current state and action.

Finally, the syncOps method returns a set of operations between entities that must be synchronized as per the sync construct. Since the Account class requires no synchronization it returns the empty set.

The Rebel library contains a restful HTTP API which derives endpoints for all the specifications and actions. These are used to trigger actions on the actors.

The translation of the MoneyTransfer class follows the same pattern. However, in this case the method syncOps does not return the empty set. It is shown in listing 4. Each SYNC action is translated to a SyncAction with a ContactPoint, which enables sending

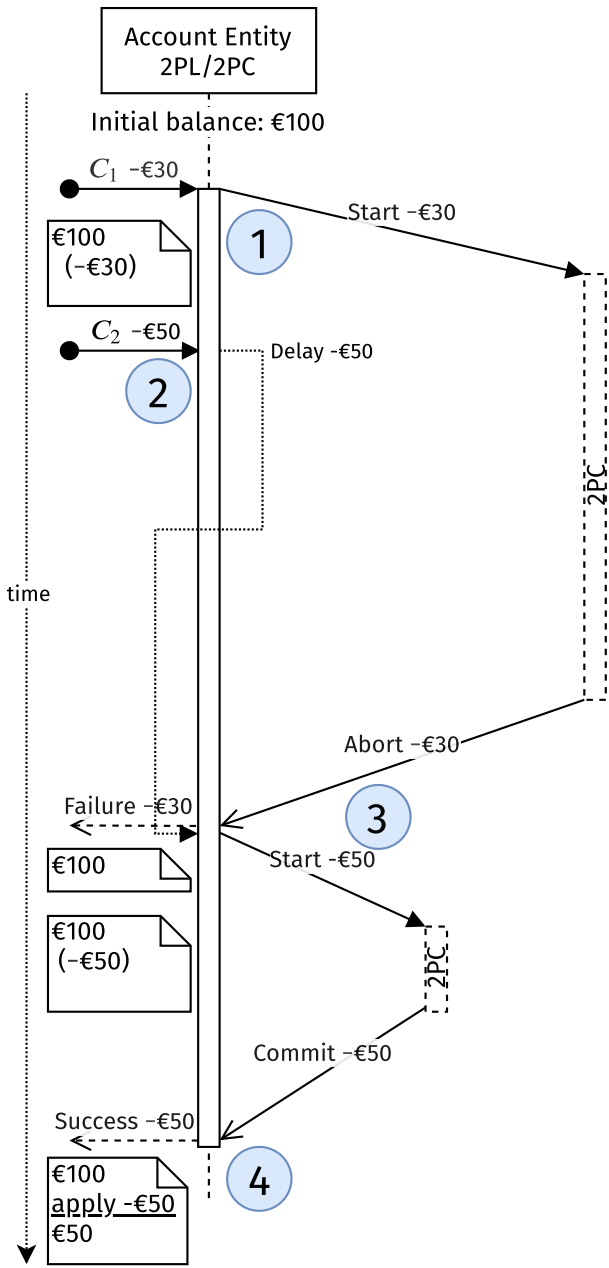
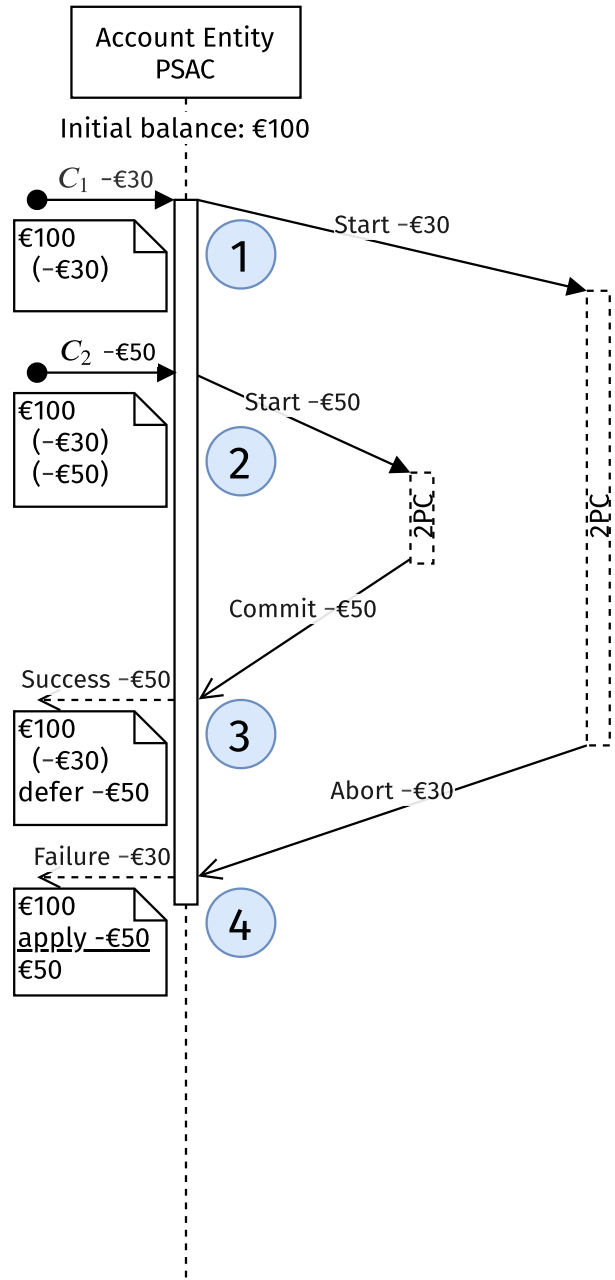

**Figure 10** Vanilla Two-Phase Commit  **Figure 11** Path-Sensitive Atomic Commit

■ **Listing 3** Generated Scala code for the Rebel Account entity (slightly simplified)

```scala
class AccountActor extends RebelFSMActor with RebelSpec[AccountState, AccountData, AccountCommand]{
 val initialState: AccountState = Init
 val allStates : Set[AccountState] = Set(Blocked, Closed, Init, Opened)
 val finalStates : Set[AccountState] = Set(Closed)

 def nextState: PartialFunction[(AccountState, AccountCommand), AccountState] = {
  case (Uninit, _: Open) => Opened
  case (Opened, _: Withdraw) => Opened
  case (Opened, _: Deposit) => Opened
  case (Opened, _: Close) => Closed
 }

 def checkPre(data: AccountData, now: DateTime): PartialFunction[AccountCommand, CheckResult] = {
  case OpenAccount(accountNumber, initialDeposit) =>
   checkPreCondition(initialDeposit ≥ EUR(50.00))
  case Close() =>
   checkPreCondition(data.get.balance.get == EUR(0.00))
  case Withdraw(amount) =>
   checkPre(amount > EUR(0.00)) combine
   checkPre(data.get.balance.get - amount ≥ EUR(0.00))
  case Deposit(amount) =>
   checkPreCondition((amount > EUR(0.00)))
 }

 def apply(data: AccountData): PartialFunction[AccountCommand, AccountData] = {
  case OpenAccount(accountNumber, initialDeposit) =>
   Initialized(AccountData(accountNumber = Some(accountNumber), balance = Some(initialDeposit)))
  case Withdraw(amount) =>
   data.map(r => r.copy(balance = r.balance.map(_ - amount)))
  case Deposit(amount) =>
   data.map(r => r.copy(balance = r.balance.map(_ + amount)))
 }

 def syncOps(data: RData): PartialFunction[AccountCommand, Set[SyncOp]] = Set.empty
}
```

messages to the sync participant living somewhere in the cluster, and the action on the sync participant itself.

## C  Detailed Rebel implementation using Akka

**FSM**  Each Rebel specification describes a single financial product. Each of these products can have multiple instances, which we call entities, that can be identified by their unique Rebel *@key*. This nicely maps to an actor definition per specification, where each running instance of this actor is an entity. Since a specification describes a state machine we piggyback on the Akka Domain Specific Language (DSL) for Finite State Machines (FSM). Akka FSM provides constructs for States, Data and Transitions. Notable features are Timeouts when no commands are received and batching of events to the persistence layer for improved performance.

◾ **Listing 4**  Generated code corresponding to the SYNC action in the MoneyTransfer

```scala
def syncOps(data: MoneyTransferData): PartialFunction[MoneyTransferCommand, Set[SyncOp]] = {
  case Book(amount, from, to) => Set(
    SyncAction(ContactPoint(Account, from), Withdraw(amount)),
    SyncAction(ContactPoint(Account, to), Deposit(amount))
  )
}
```

**Cluster**  The Akka cluster feature allows us to run our application on multiple servers, by supplying a mechanism to add extra nodes to the Akka cluster and location-transparently send messages to actors on remote cluster nodes. Akka takes care of the setting up of the cluster and and the joining and leaving of nodes. This is what allows our application to scale horizontally in the number of nodes and therefore total the number of running actors and the amount system resources.

**Sharding**  Rebel specifications allow interaction with other specification in pre-, post-conditions and synchronised actions. Other entities can be accessed by using the specification name and identity ($@key$), e.g. $Account[this.from]$. The identity referenced, can be from a specification field or event field.

Akka Sharding allows us to distribute the running actors over the cluster nodes. Each cluster node can start a shard region, which is used to send messages to a certain type of actor somewhere in the cluster. An actor can be reached by a unique logical identifier. The sharding feature makes sure that for each unique identifier only one actor is active in the whole cluster. If a message is send to an identity that is not yet running, the corresponding actor will be started somewhere in the sharding cluster.

Rebel identity nicely maps to the logical identity of Akka Sharding. In the target application we use a sharding region per specification. This allows us to send messages to each individual actor, without knowing or caring on which node it is running in the actual Actor Cluster. If the entity is not yet created, Sharding will make sure it is started and usable. This is called *location transparency*.

**Persistence**  Sharding allows us to distribute the actors over the system and makes sure only once actor is running per entity. In order to be able to durably store the data and state of an entity we use Akka Persistence.

Akka Persistence is based on Event Sourcing (ES) and Command Query Responsibility Segregation (CQRS). This means that for each event that a Rebel entity can process, a Command and an Event is defined. Respectively denoting the intention to let the event happen and the immutable proof that the event occurred.

Event sourcing means that we create an immutable log of all sequential immutable events that happened.

In our application this means that for each actor corresponding with a Rebel entity we store the incoming command after a precondition check in our persistency layer[4] as an intention to execute this command. After successfully persisted, the command is executed. If valid an event of the transition will be committed to the persistence layer and side effects to the internal state will be executed. Akka Persistence makes sure other incomming commands are delayed until the in-progress command is handled completely.

The result of recording all the events in a persisted log is that we can restart an actor and replay all the events that happened and get it back into the last committed state. Because Sharding makes sure there is only one actor with the same logical identity running at a single moment in time, we can be sure that only a single persistent actor is writing to the log and know that its internal entity level state is consistent.[5]

In the event of an actor or cluster node crashes, the entity can be restarted on another node without loss of data. This also means that the persistency guarantees are heavily dependent on the guarantees of the underlying persistence journal implementation.

As persistence backend we use Cassandra 3. This is a production-ready backend for Akka Persistence and also the mostly used.

**HTTP** The Rebel events are exposed as REST endpoint for each logical identifier on which commands can be triggered to the corresponding actors.[6] This uses Akka HTTP for non-blocking IO and can be automatically derived from the available generated specification and event implementations. We use the *circe* JSON library to automatically derive JSON encoders and decoders for each of the events, based on the generated case classes. This means that for the entire REST interface almost no additional code has to be generated, next to the field and event definitions. A Scala worksheet file is available to manually generate example JSON-documents that the system accepts.

An Open API specification[42] definition is generated which corresponds to the generated REST interface. This allows for easy consumption of the interface.

These endpoints are used for the experiment by the load generator.

---

[4] We use Cassandra, but many more journaling plugins are available. The queries are configured to write and read with Quorum, so we know that each command is persisted safely before it is handled.

[5] Also because the actor only handles a single message and therefore a single command at the same time

[6] url template: POST /SPECIFICATION-NAME/:ID/EVENT-NAME

## About the authors

**Tim Soethout** is a senior software developer at ING Bank. As guest researcher in the Software Analysis and Transformation group at Centrum Wiskunde & Informatica (CWI) he works on his PhD-research on leveraging domain models to improve performance and scalability of distributed enterprise applications. He can be reached at Tim.Soethout@ing.com.

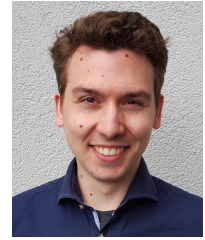

**Tijs van der Storm** is senior researcher in the Software Analysis and Transformation group at Centrum Wiskunde & Informatica (CWI), and full professor in Software Engineering at the University of Groningen. His research focuses on improving programmer experience through new and better software languages and developing the tools and techniques to engineer them in a modular and interactive fashion. For more information, see http://www.cwi.nl/~storm. He can be reached at T.van.der.Storm@cwi.nl.

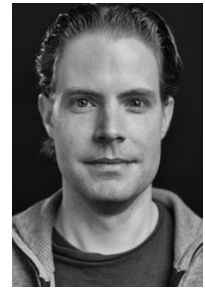

**Jurgen J. Vinju** is full professor of Automated Software Analysis at Eindhoven University of Technology, research group leader at Centrum Wiskunde & Informatica (CWI), and senior language engineer and co-founder of SWAT.engineering. He studies the design and evaluation of (applications of) meta programming systems to get the complexity of source code maintenance under control. Examples are metrics and analyses for quality control or debugging, and model driven engineering for code generation. For more information, see http://www.cwi.nl/~jurgenv. He can be reached at Jurgen.Vinju@cwi.nl.

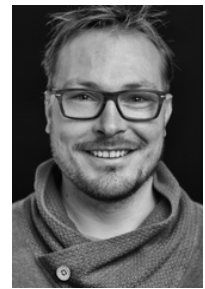